\documentclass[12pt]{article}  

\usepackage{amsmath,amsfonts,amssymb}
\usepackage{graphicx}
\usepackage{float}
\usepackage[colorlinks=true, allcolors=blue]{hyperref}

\providecommand{\keywords}[1]
{
  \small	
  \textbf{\textit{Keywords---}} #1
}

\title{Current status of PAPYRUS : the pyramid based adaptive optics system at LAM/OHP}

\author{
{Muslimov E.}$^{c}$,{Levraud N.}$^{a,b}$,{Chambouleyron V.}$^{a,b}$ ,\\
{Boudjema I.}$^{a}$ ,{Lau A.}$^{a}$, {Caillat A.}$^{a}$,{Pedreros F.}$^{a}$,\\
{Otten G.}$^{a}$,{El Hadi K.}$^{a}$, {Joaquina K.}$^{a}$,{Lopez M.}$^{a}$,\\
{El Morsy M.}$^{a}$,{Beltramo Martin O.}$^{a}$,{Fétick R.}$^{a,b}$,\\
{Ke Z.}$^{a}$,{Sauvage J-F.}$^{a,b}$, {Neichel B.}$^{a}$, {Fusco T.}$^{a,b}$,\\ {Schmitt J.}$^{d}$, {Le Van Suu A.}$^{d}$,{Charton J.}$^{e}$,{Schimpf A.}$^{e}$,\\
{Martin B.}$^{e}$,{Dintrono F.}$^{e}$, {Esposito S.}$^{f}$,{Pina E.}$^{f}$\\
\small a--{Laboratoire d'Astrophysique de Marseille, France}\\
\small b--{DOTA,ONERA, Université Paris Saclay, F-91123 Palaiseau, France}\\
\small c--{NOVA Optical IR Instrumentation Group, Dwingeloo, Netherlands}\\
\small d --{Observatoire de Haute Provence, Saint-Michel-L'observatoire, France}\\
\small e --{ALPAO, Montbonnot-Saint-Martin, France}\\
\small f --{Osservatorio Astrofisico di Arcetri, INAF, Italie}\\
}

\begin{document} 
\maketitle

\begin{abstract}
The Provence Adaptive optics Pyramid Run System (PAPYRUS) is a pyramid-based Adaptive Optics (AO) system that will be installed at the Coude focus of the 1.52m telescope (T152) at the Observatoire de Haute Provence (OHP). The project is being developed by PhD students and Postdocs across France with support from staff members consolidating the existing expertise and hardware into an R\&D testbed. This testbed allows us to run various pyramid wavefront sensing (WFS) control algorithms on-sky and experiment on new concepts for wavefront control with additional benefit from the high number of available nights at this telescope. It will also function as a teaching tool for students during the planned AO summer school at OHP. To our knowledge, this is one of the first pedagogic pyramid-based AO systems on-sky. The key components of PAPYRUS are a $17\times17$ actuators Alpao deformable mirror with a Alpao RTC, a very low noise camera OCAM2k, and a 4-faces glass pyramid. PAPYRUS is designed in order to be a simple and modular system to explore wavefront control with a pyramid WFS on sky. We present an overview of PAPYRUS, a description of the opto-mechanical design and the current status of the project.
\end{abstract}

\keywords{Adaptive Optics, Pyramid sensor}

\section{INTRODUCTION}
\label{sec:intro}  

\subsection{Presentation of the project}

The Provence Adaptive-optics PYramid RUn System (PAPYRUS) is an adaptive optics system to be installed
at the Coude focus of the T152 telescope (diameter of 1.52 m) at Observatoire de Haute Provence (OHP).
The main specificity of Papyrus is to use a pyramid wavefront sensor, which shows better SNR
performances than the usual Shack-Hartmann wavefront sensor. The pyramid wavefront sensor has
widely been described theoretically \cite{fauvarque2017,fauvarque2019kernel} and tested on bench \cite{janin2019adaptive}. PAPYRUS is a demonstrator of feasibility on sky
and a TRL maturation step in between the current tests performed at LAM and the future generation of
pyramid wavefront sensors such as ELT/HARMONI.\\

The specificity of PAPYRUS is to be a low cost and student led project. Indeed PAPYRUS will use
components already available at LAM and ONERA in order to drastically reduce hardware costs. The
PAPYRUS project encompasses all the following steps:
\begin{itemize}
    \item System specification and requirements
    \item System optical and mechanical design
    \item Assembly and testing at LAM
    \item Installation and testing at OHP
    \item Maintenance and eventual upgrades at OHP
\end{itemize}

\subsection{PAPYRUS project objectives}

As stated above, PAPYRUS is a demonstrator of capability to place a pyramid wavefront sensor on
sky. The LAM/GRD and ONERA teams will gain a valuable knowledge on the pyramid behaviour,
especially regarding the development of specific control laws. The impact of the PSF modulation
on the pyramid will be a major topic of study thanks to a dedicated modulation mirror and to a dedicated
pyramid focal plane camera \cite{chambouleyron2021focal} (“gain scheduling camera”). Management of the pyramid optical gains and non-linearities is still a subject under active research. PAPYRUS might be upgraded in the future to test new techniques or components for astronomy with AO.\\

Moreover, PAPYRUS data will be used to check image post-processing possibilities from pyramid AO systems. PSF analysis, PSF estimation \cite{fetick2020blind} or deconvolution \cite{mugnier2004mistral} is envisaged.\\

Finally, PAPYRUS will be a pedagogical on-sky AO bench located near Marseille (France) to teach on AO techniques students coming from different laboratories or summer schools at OHP. Observing proposals may be issued by astronomers once the system performances and sky coverage have been validated on sky.


\subsection{The Observatoire de Haute Provence (OHP) site}

The astronomical site of OHP, located inthe South-East of France, has been chosen for accessibility reasons. It is the closest observatory to the development laboratory at LAM, the OHP T152 telescope has available space for installing an AO bench and available observation time. The telescope observation time will have to be shared with the AURELIE instrument, the latter being used only 50 (clear) nights per year approximately. The OHP suffers from approximately 60-80 cloudy nights per year (figure 3), it remains around 220-240 available clear nights per year for PAPYRUS.\\

The OHP is supporting the PAPYRUS effort. OHP will allow installation of the AO bench on the telescope and teaching some PAPYRUS people to use the telescope in autonomy. The T152 telescope is made of four mirrors of 88\% reflectivity each, providing a total transmission of 52\%. Recent coating on the primary mirror might slightly improve this transmission factor. The main drawback is the T152 important pointing error of $\pm 2$~arcmin. Once the star has been located in the T152 within this error range, a manual search would have to be performed to make it appear on the PAPYRUS camera. This manual searching phase will reduce slightly the effective AO lock time of PAPYRUS but is perfectly manageable according to OHP astronomers and our own usage of the telescope.

\section{Adaptive Optics Design}

\subsection{Atmospheric conditions at OHP}

\begin{figure}[h]
\centering
	\includegraphics[width=0.6\columnwidth]{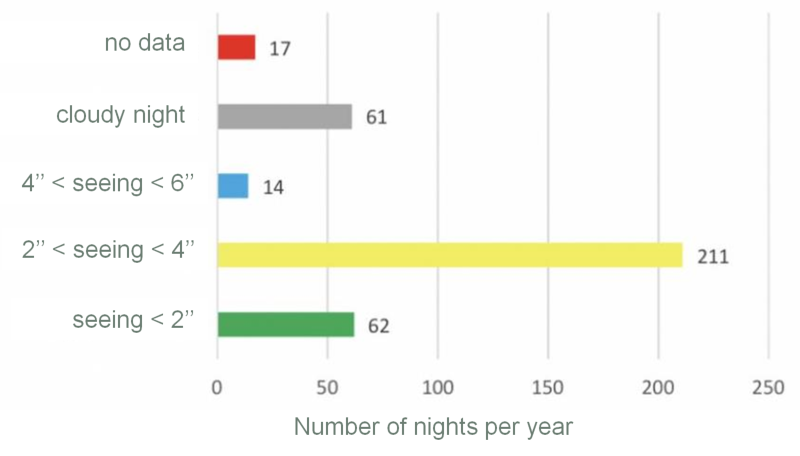}
    \caption{Statistics of seeing at OHP in 2018. Image taken and adpted from OHP website.}
    \label{fig:ohpseeing}
\end{figure}

The adaptive optics design and performances depends on the atmospheric conditions at the observing site. Data from OHP \cite{OHPwebsite} in Fig.\ref{fig:ohpseeing} show $\sim 60$ nights of good seeing $s<2"$. The median of the data is in between $s=2"$ and $s=4"$, representing approximately $200$ nights per year. These values give a typical Fried parameter $4<r_0<6$ cm at a wavelength of $600$ nm.\\

The median wind speed at OHP is about $5$ m/s, with bursts up to $7$ m/s.

\subsection {Adaptive optics error budget}

The performances of the system may be stated in term of Strehl ratio between the AO corrected long exposure and the T152 diffraction limit. Using the Marechal approximation, this Strehl ratio is related to the residual electromagnetic phase variance over the pupil as
\begin{equation}
    S_r = e^{-\sigma^2}
\end{equation}
The usual way of balancing an AO error budget\cite{rigaut1998} is to write the phase variance $\sigma^2$ as the quadratic sum of different contributions
\begin{equation}
    \sigma^2 = \sigma_\text{fitting}^2 + \sigma_\text{aliasing}^2 + \sigma_\text{noise}^2 + \sigma_\text{temporal}^2 + \sigma_\text{others}^2
\end{equation}
The definition and computations of these terms is given below. These expressions being chromatic, we choose to give all values in the visible, at a wavelength of $635$ nm. 

\subsubsection{Fitting error}

The so-called fitting error is due to the limited number of actuators of the deformable mirror. Thus the high spatial frequencies of the turbulent phase cannot be corrected by the AO system and are present in the residuals. The expression of this error is
\begin{equation}
    \sigma_\text{fitting}^2 = 0.27 \left(\frac{d}{r_0}\right)^{5/3}
\end{equation}
with $d$ the pitch between two actuators and $r_0$ the Fried parameter of the turbulence. For our $17 \times 17$ deformable mirror on a $D=1.52$ m pupil, the pitch is $9.5$ cm, giving an error of
\begin{equation}
    \sigma_\text{fitting}^2 = 0.6 \text{\;rad}^2
\end{equation}

\subsubsection{Aliasing error}

The wavefront measurement is sampled for detection, producing aliasing of the high frequencies onto the low frequencies. This information from the high frequencies enters in the AO loop and generates a deformable mirror response. However the aliasing is smaller on a pyramid wavefront sensor than ona Shack-Hartmann wavefront sensor. It is then neglected in our AO budget
\begin{equation}
    \sigma_\text{aliasing}^2 = 0 \text{\;rad}^2
\end{equation}

\subsubsection{Noise error}

The photon noise and detector noise both produce intensity fluctuations on the wavefront sensor and consequently in the estimated wavefront. The advantage of our AO design is to use a low-noise OCAM$^2$K detector for wavefront sensing. Detector noise is thus neglected.\\

Since PAPYRUS will be targeting bright stars in its early design, photon noise will be mitigated. In order to achieve a correct SNR $=10$ on each measurement, one needs $100$ photons per point of measure per frame. This flux on the detector translates into flux of the AO guiding target as
\begin{equation}
    N_\text{ph/m2/s} = N_\text{ph/meas/frame}\frac{F}{\eta}\left(\frac{n_{meas}}{D}\right)^2
\end{equation}
where $N_\text{ph/m2/s}$ is the flux of the guiding target in photons per m$^2$ per second, $N_\text{ph/meas/frame}=100$ is the desired flux on the detector in photons per point of measurement per frame, $F=300$ Hz is the frame frequency, $\eta\simeq 10\%$ is the global photon efficiency from telescope to the WFS detector, $n_{meas}=16$ is the number of measurements in the diameter, $D=1.52$ m is the telescope diameter. One finds
\begin{equation}
    N_\text{ph/m2/s} = 33\times 10^6
\end{equation}
For a wavelength of $700$ nm and a filter width of $20$ nm, this flux is obtained for stars of magnitude $V\leq 4$. With this condition on the star magnitude the noise error propagated in the loop is neglected
\begin{equation}
    \sigma_\text{noise}^2 = 0 \text{\;rad}^2
\end{equation}

\subsubsection{Temporal error}

The temporal error comes from the delay between the actual phase present on the pupil and its correction by the deformable mirror. The temporal error writes
\begin{equation}
    \sigma_\text{temporal}^2 = 0.04 \left(\frac{V}{DB}\right)^2 \left(\frac{D}{r_0}\right)^{5/3} \sum_{n=1}^{N_r}(n+1)^{-2/3}
\end{equation}
with $V$ the turbulence equivalent wind speed (Taylor frozen flow hypothesis), $B$ the AO loop bandwidth and $N_r$ the number of radial modes corrected. For a control law including an integrator, the equivalent bandwidth writes
\begin{equation}
    B = \frac{F}{2\pi}\sqrt{\frac{g}{1+2\tau F}}
\end{equation}
with $g$ the integrator gain and $\tau$ the loop delay. A reasonable choice of gain ensuring both performances and robustness is
\begin{equation}
    g = \frac{1}{1+\tau F}
\end{equation}
Considering $\tau F = 2$ frames delay for PAPYRUS and a wind speed $V = 7$ m/s, we deduce the bandwidth and then the temporal error
\begin{equation}
    \sigma_\text{temporal}^2 = 5.9 \text{\;rad}^2
\end{equation}

\subsubsection{Other sources of error}

Other sources of error include, but are not limited to, scintillation on the pupil, non-common path errors, and calibration errors. These errors are not taken into account, especially in front of the important temporal error.

\subsection{Conclusions on the budget}

The error budget above is highly dominated by the temporal error. This error is established with the highest wind speed at OHP, and thus corresponds to a worst case of PAPYRUS usage. It is planned to mitigate this error by implementing more complex control laws to increase the bandwidth, and using a better RTC. Indeed we plan to replace our current computer (run on Matlab) by a dedicated ALPAO high-speed RTC.\\

\begin{figure}[h]
\centering
	\includegraphics[width=0.7\columnwidth]{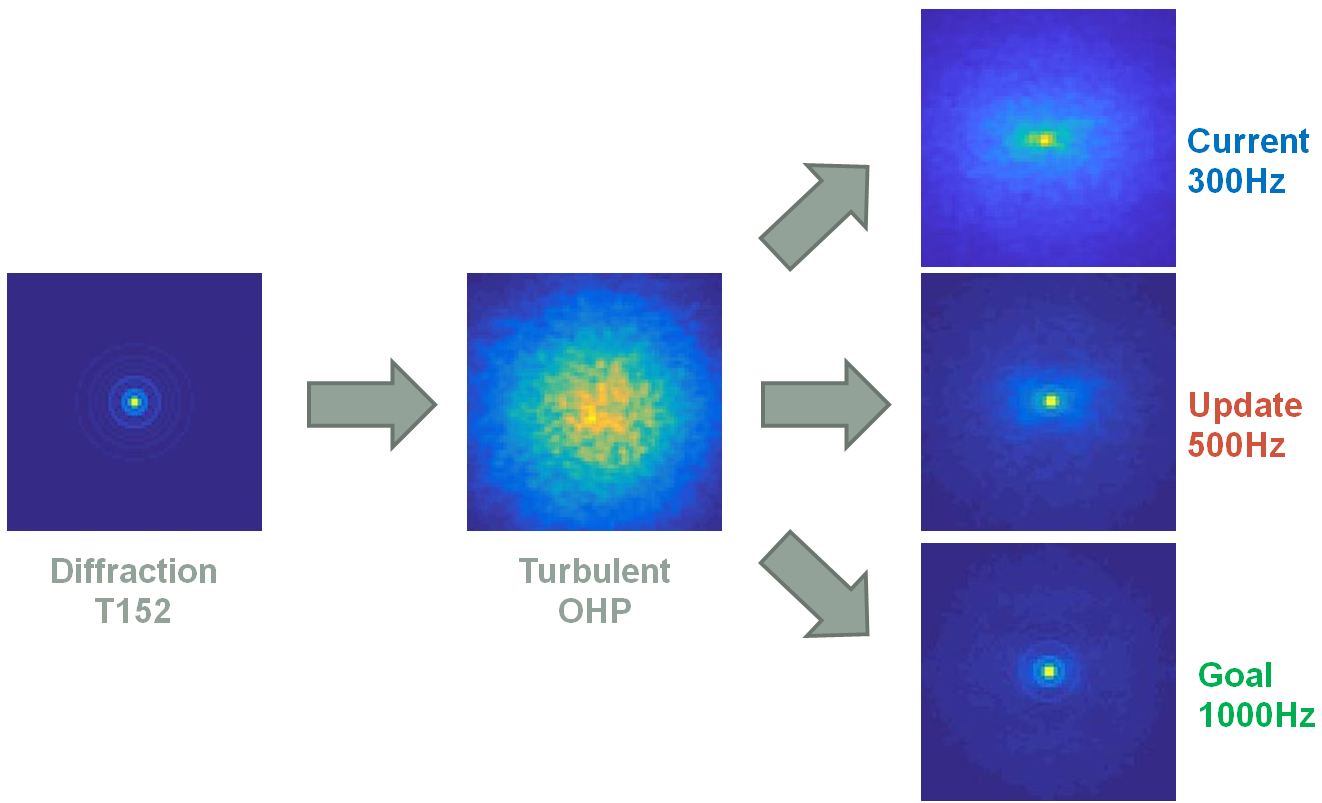}
    \caption{PSF from end-to-end simulations. Three cases of loop frequency are considered to simulate the PAPYRUS PSF.}
    \label{fig:oomaopsf}
\end{figure}

Adaptive-optics corrected PSF are simulated using OOMAO\cite{conan2014} end-to-end simulations with the PAPYRUS system parameters, see Fig. \ref{fig:oomaopsf}. The temporal error being the dominant one, we use three different AO loop frequencies for the simulations. It shows that the current $F=300$ Hz is quite limitating and incline us to go for a better RTC at $F=500$ Hz at least.

\section{Optical Design}

\subsection{Initial data and design challenges}
The optical design is built to work together with the T152 telescope at OHP. It represents an \textit{F/28.56} Cassegrain system with the focal length of 43.4 m and the maximum field of view (FoV) of $\pm 3^{\circ}$.
The telescope has residual aberrations, which don't exceed $16.8 \mu m$,while the Airy disk radius is $13.95 \mu m$. 

Also, we had certain cameras available for the scientific and wavefront sensing branches as well as the commercial deformable and tip-tilt mirrors. Their parameters were also driving the design.

Thus, except of the telescope parameters the initial data for the optical design development included:
\begin{enumerate}
    \item Diameter of off-the-shelf mirror used for slow tip/tilt =25.4 mm;
    \item Diameter of the existing DM = 37.5 mm;
    \item Max diameter of the existing fast tip/tilt mirror = 12 mm;
    \item Format of ORCA science camera = $13.312 \times13.312 mm^2$ with $2048 \times 2048$ pixels;
    \item Format of OCAM WFS camera= $5.76 \times 5.7 mm^2$ with $240\times240$ pixels;
    \item The measured  parameters of existing glass pyramid: facet angle = $8.9^{\circ}$, material LF5, leading to deflection angle of $\pm 5.44^{\circ}$;
    \item Science camera FoV = $\pm 1"$;
    \item WFS FoV = $\pm 0.25:$;
    \item Working wavebands =  400-1000 nm (Science camera) and  630-690 nm (WFS with filter).
\end{enumerate}

In addition to these initial data there were some limitations, which made the development challenging. First, the project budget and the development time were limited. Therefore we had to rely only on off-the-shelf components and iteratively repeat the design to find compromise solutions. Second, since our demonstrator is targeting multiple tasks, its' design should be flexible with a possibility to replace some components,  to connect testing and alignment modules or to feed other instruments through optical ports, to get an access to any par of it for educational purposes.  Third, the optical design of the bench should provide a high image quality to minimise the instrumental WFE and the difference between the science and WFS branches. Finally, there were limitations in overall dimension and optical interfaces because the bench will be installed 
on a telescope among the existing instruments.

\subsection{Optical design overview and analysis}

\begin{figure}[H]
\centering
	\includegraphics[width=10 cm]{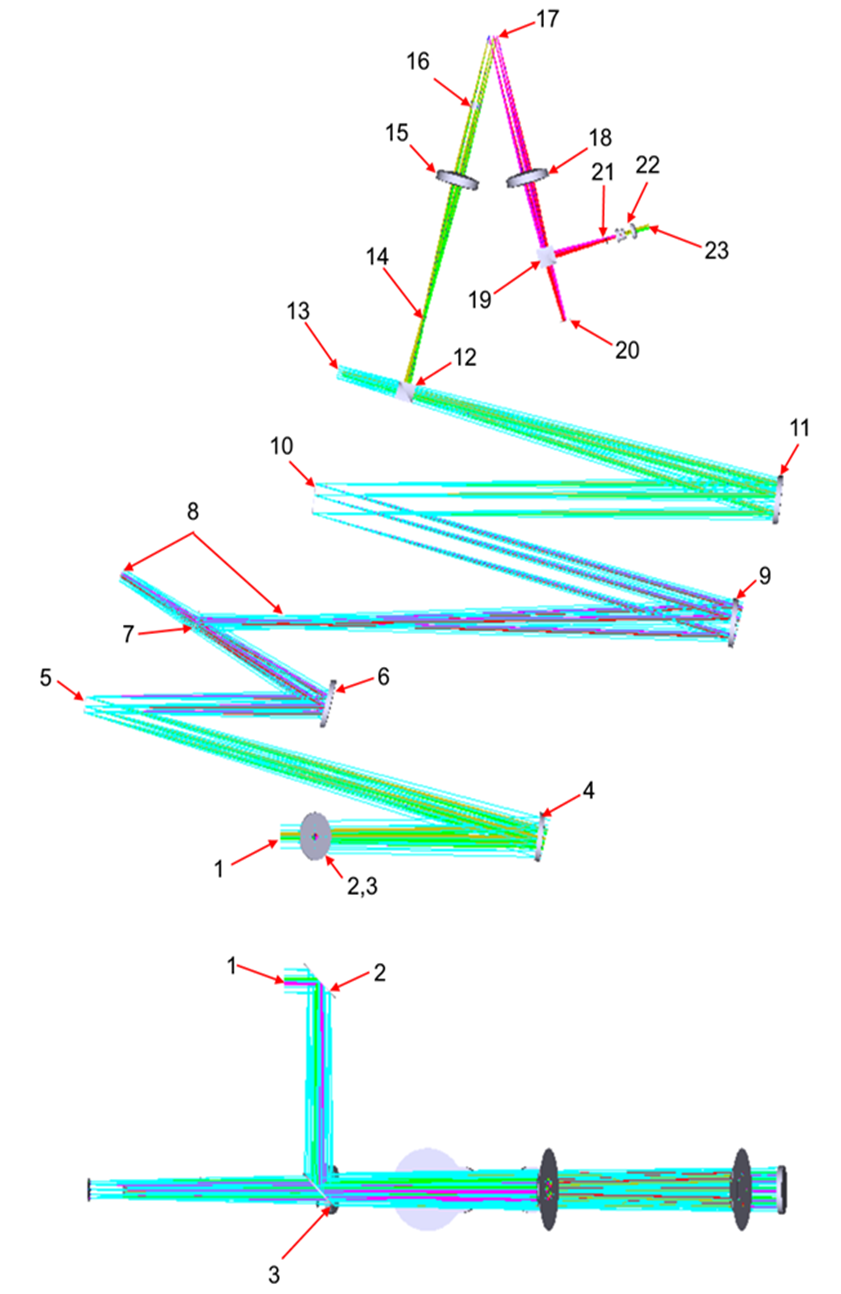}
    \caption{General view of the PAPYRUS optical design  (see the text for the elements notation). }
    \label{fig:layout}
\end{figure}

With all of the above-listed initial values and limitations we developed the following optical design (see Figure~\ref{fig:layout}).
In order to provide flexible connection with the telescope and compensate the height difference the movable periscope unit 2, 3 is used to pick up the beam from the  telescope focal plane 1. Hereafter we use two optical relays to form intermediate focal and pupil planes and place key components there. The relays are built on the basis of off-axis parabolic (OAP) mirrors. use of OAP allows to exclude chrmatic aberrations and significantly reduce the overall dimensions, although makes the alignment more difficult. The beam is collimated (4) and the pupil is imaged on the slow tip-tilt mirror 5. Further the beam is focused again (6) with $-0.5^x$ magnification and the pupil is re-imaged to the infinity. Folding mirror 7 represents an optical port, which can be used to couple the telescope simulator optical to the intermediate focus 8. The beam is collimated again (9) and the pupil is imaged onto DM 10. Then it is focused (11) in the same way with $–1^x$ magnification and the pupil projection to the infinity. The image formed in focal plane is detected by science camera 13. Before the science camera a beamsplitter 12 is mounted to feed the WFS branch. Note, that the WFS branch uses lens optics, because the components there should have short focal lenghts and work with relatively large angular fields. The WFS $1^{st}$ intermediate focus 14 can be used to set a diaphragm. The beam is re-collimated (15) and the pupil is projected to the modulation mirror 17. In the formed parallel beam after 15 a filter 16 can be mounted to minimize the WFS chromatism. It should be noted that some chromatic aberration is inherent to the pyramid itself regardless of the auxiliary optics properties. The beam is focused (18), while the pupil is projected to the infinity. The focused beam is split (19) to form two identical images at the tracking camera 20 and the tip of pyramid 21. The pyramid splits the beam in quadrants with $5.44^\circ$ deflection. The resultant beams are collimated (22) and the pupil is projected to the WFS camera 23. 

Thus, the optical design consists of a series of functional elements, each served with a collimating and focusing relay. Since each focusing component should project the pupil to the infinity, the pupils coincide with the focusing components focal planes. For the same reason, the collimating components form the pupil images in their back focal planes.
The actual focal lengths and magnifications in the system were choosen to fit the initial design conditions and to obtain the focal lengths, $F/\#$ and field values close to those of available commercial components.\\

\begin{figure}[H]
\centering
	\includegraphics[width=16 cm]{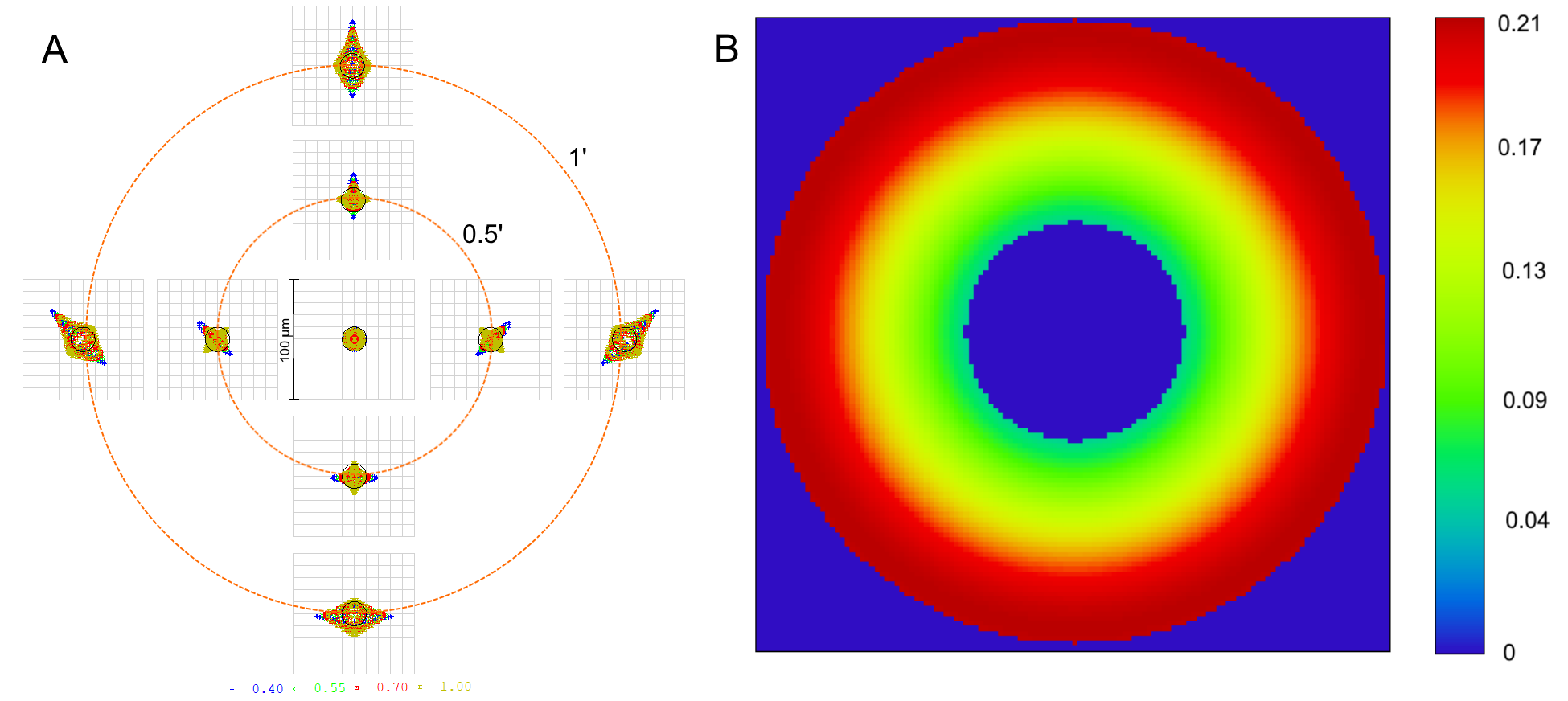}
    \caption{Image quality at the scientific camera: A -- spot diagrams over the FoV (the colors correspond to wavelengths in $\mu m$); B -- wavefront error in waves $@ 0.7 \mu m$  referred to the exit pupil plane. }
    \label{fig:sci cam}
\end{figure}

The Figure \ref{fig:sci cam} shows the image quality in the field of view of science camera. The spot diagram root mean square (RMS) radii vary between 5.3 and 14.9 $\mu m$, while the Airy disk radius is $10.1 \mu m$. So, the image quality is diffraction limited on-axis and slightly degrades towards the field edge. The WFE RMS value is $0.048 \lambda$ with the maximum of $0.16 \lambda$ at 700 nm. The on sky quality will  however be dominated by AO residuals and anisoplanatism, so this design is completely accepted for PAPYRUS.

\begin{figure}[H]
\centering
	\includegraphics[width=16 cm]{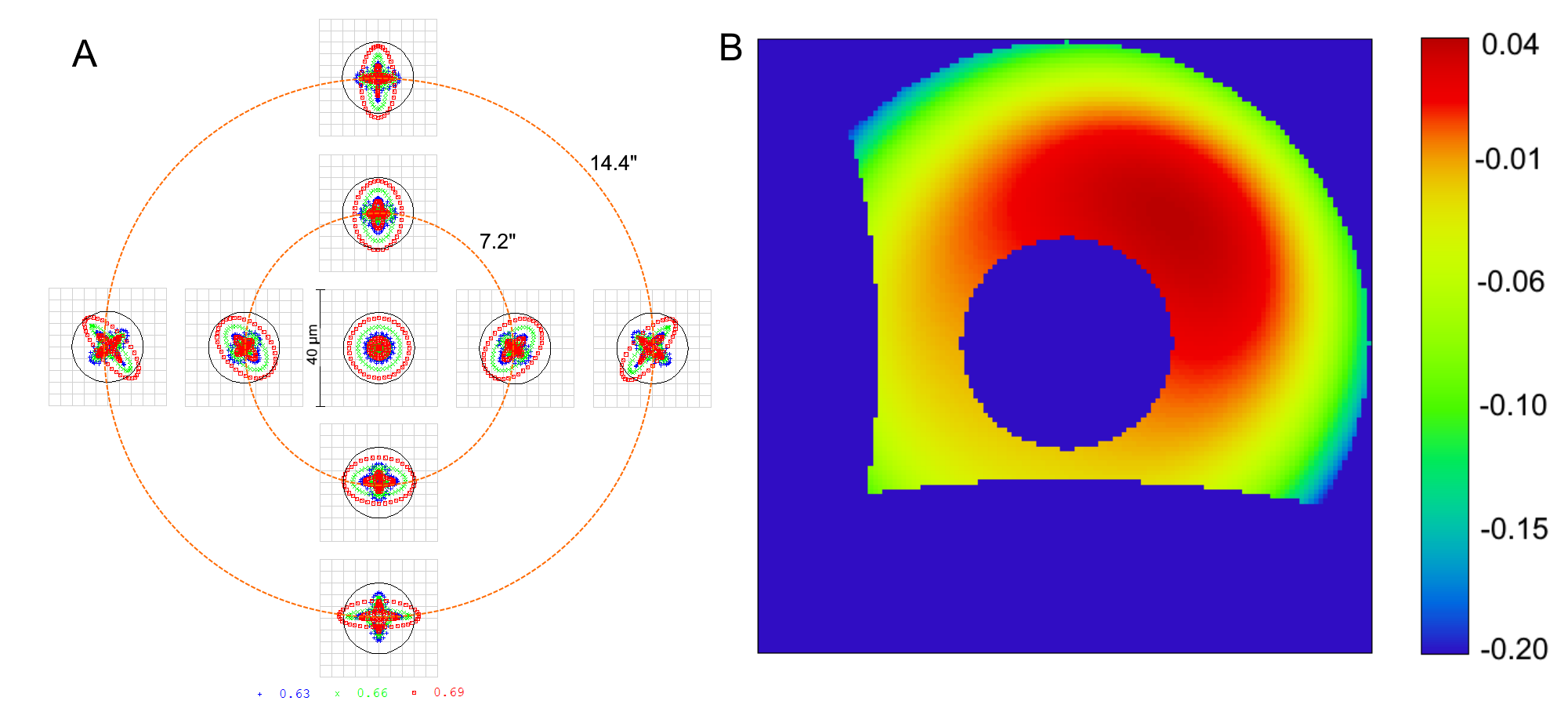}
    \caption{WFS branch image quality at the tracking camera: A -- spot diagrams over the FoV (the colors correspond to wavelengths in $\mu m$ with filter); B -- wavefront error in waves $@ 0.7 \mu m$  referred to the exit pupil plane. }
    \label{fig:track cam}
\end{figure}

The image quality assessment for the WFS branch is shown in Figure~\ref{fig:track cam}. The geometrical spot size remains stable across the field, since the nominal pyramid FoV is only $\pm14.4"$. It has the RMS radius of 5.1-6.0 $\mu m$ , while the Airy radius is 12.1 $\mu m$. The WFE for a single pyramid facet is $0.046 \lambda$ RMS at 700 nm. It becomes possible to reach the diffraction limit only with use of a red band filter. Otherwise the chromatic aberrations inherent to this system would blur the spot images by factor of 4. 

\begin{figure}[H]
\centering
	\includegraphics[width=16 cm]{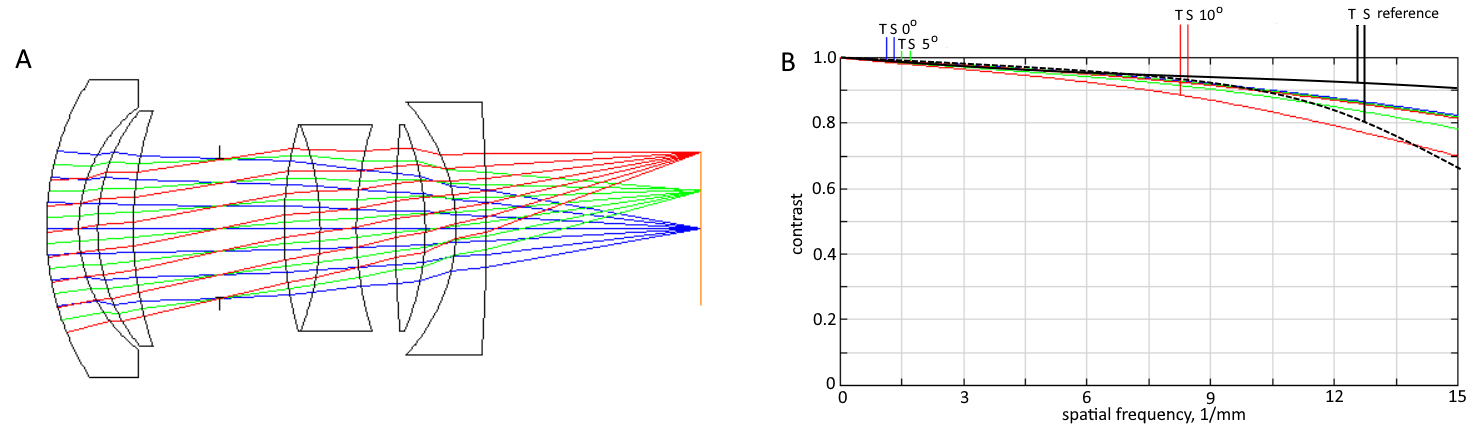}
    \caption{Reconstructed design of the WFS lens: A -- general view, B -- MTF in comparison with the reference curve from open sources.}
    \label{fig:canon}
\end{figure}

We would like to note separately the design issue related to the last WFS lens, which forms the image on the camera after the pyramid. It should work with a relatively low $F/\#$ and extended field of view, but also in an unordinary setup, where the pupil is shifted and doesn't coincide with the nominal aperture stop or its' images. In the meantime this lens should be an off-the-shelf component and provide a high image quality. A $Canon^{TM}$ EF-S 24mm F/2.8 STM pancake photographic lens was chosen, since it is has small total length and provide high image quality. However, to account for possible vignetting and aberrations growth is this setup we had to reconstruct an approximate optical design of the lens. An approximate algorithm of this reverse-engineering is given below. We hope that it can be useful for those, who are going to use commercial lenses in non-standard laboratory setups. 
\begin{enumerate}
    \item The general data about the lens like the number of lenses and groups as well as the composition of positive and negative components were taken from the manufacturer\cite{CANON}.
    \item The closest patent was found\cite{Ohtake97}. The lens composition is similar to the known one and the publication date is the closest ot the lens commercial release. 
    \item The patent lens was scaled to fit the focal length and the first lens diameter of the commercial lens.
    \item The lens was optimized to fit such criteria as the MTF values and vignetting known from open sources. The radii of curvature and thicknesses were used as free variables. In addition to standard boundary conditions defining the edge and center thicknesses, the overall system length and diameters and optical powers of individual components were restricted. The lens design after optimization is shown in Figure~\ref{fig:canon},
\end{enumerate}

\begin{figure}[H]
\centering
	\includegraphics[width=\columnwidth]{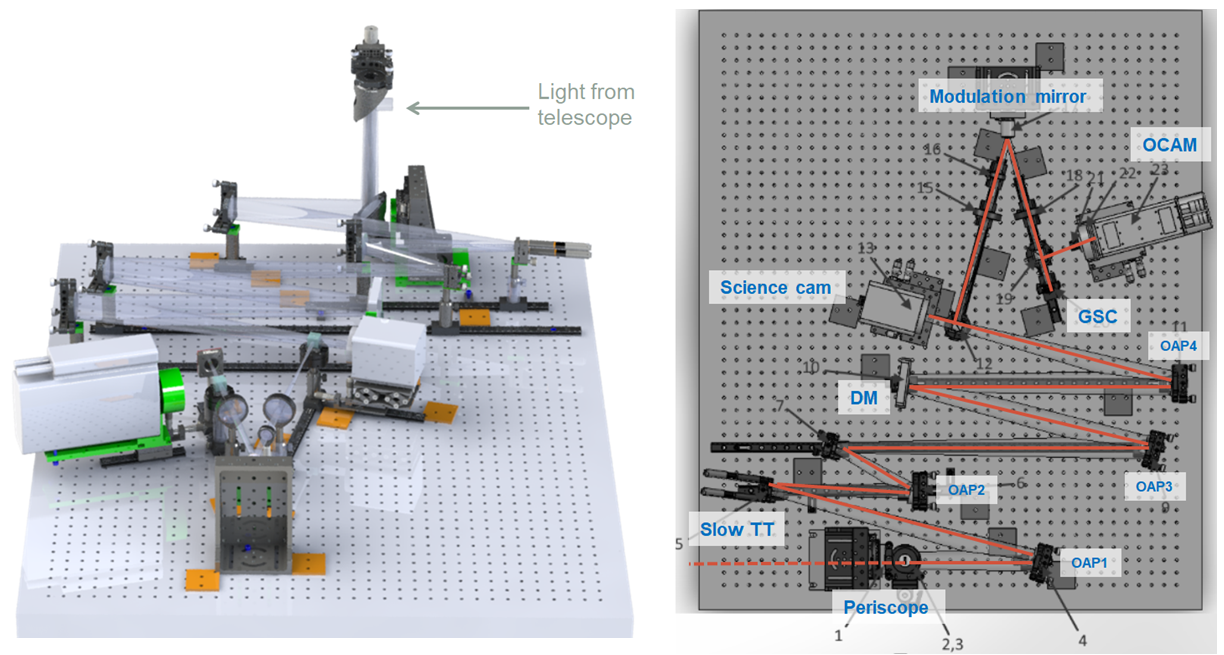}
    \caption{Opto-mechanical model of PAPYRUS in 3D view (left) and top view (right). GSC: gain scheduling camera. DM: deformable mirror. OAP: off-axis parabola. Slow TT: slow tip-tilt mirror.}
    \label{fig:papyrusview}
\end{figure}

The figure \ref{fig:papyrusview} shows the opto-mechanical implementation of PAPYRUS with the different elements already presented before. The optomechanical design uses off-the-shelf parts for all the adjustments and holding the components and has as less as possible custom parts only as mechanical interface adapters. The mechnaical mounts provide the necessary number of degrees of freedom for every unit. The entire bench is assembled on a single breadboard. 
The optical path is divided into straight segments and all the components of such a segment are mounted on a single optical rail. Position of each of the optical rails is given by a couple (line+corner) flat templates. The periscope assembly and the cameras are mounted via custom adapters, which can be 3d-printed or milled in aluminium. All the OAP mirrors have adapters to compensate the vertex lateral shift. Finally, the WFS camera together with the pyramid are mounted in a custom housing with adjustment rings and attached directly to the WFS camera C-mount.   
This optomechanical implementation allows some room for future upgades of the AO bench, and visiting scientific instrument to be adapted to the bench.

\section{Current Status of papyrus}

\subsection{Assembly Integration and Tests}

The AIT of PAPYRUS will be organized following this strategy: 
\begin{itemize}
    \item component validation at LAM
    \item global bench alignement and validation at LAM
    \item final coupling and commissioning at T152 OHP. 
\end{itemize}

\subsubsection{Test of the pyramid}

\begin{figure}[H]
\centering
	\includegraphics[width=0.5\columnwidth]{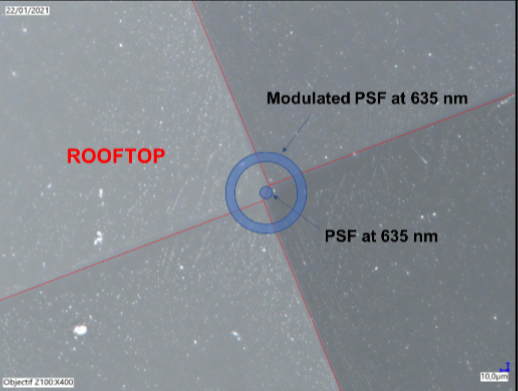}
    \caption{Picture of the PAPYRUS pyramid tip, showing a "rooftop" configuration.}
    \label{fig:pyramid}
\end{figure}

The critical components of PAPYRUS have been tested individually. The glass pyramid is obviously one of these critical components. Pictures of the pyramid tip (Fig. \ref{fig:pyramid}) show a "rooftop" configuration due to manufacture precision. The rooftop size is approximately the diffraction size that would prevent WFS sensing in one direction if non modulated. However the modulation of the PSF around the pyramid tip solves this issue.

\subsection{Testing the WFS branch}

\begin{figure}[H]
\centering
	\includegraphics[width=0.9\columnwidth]{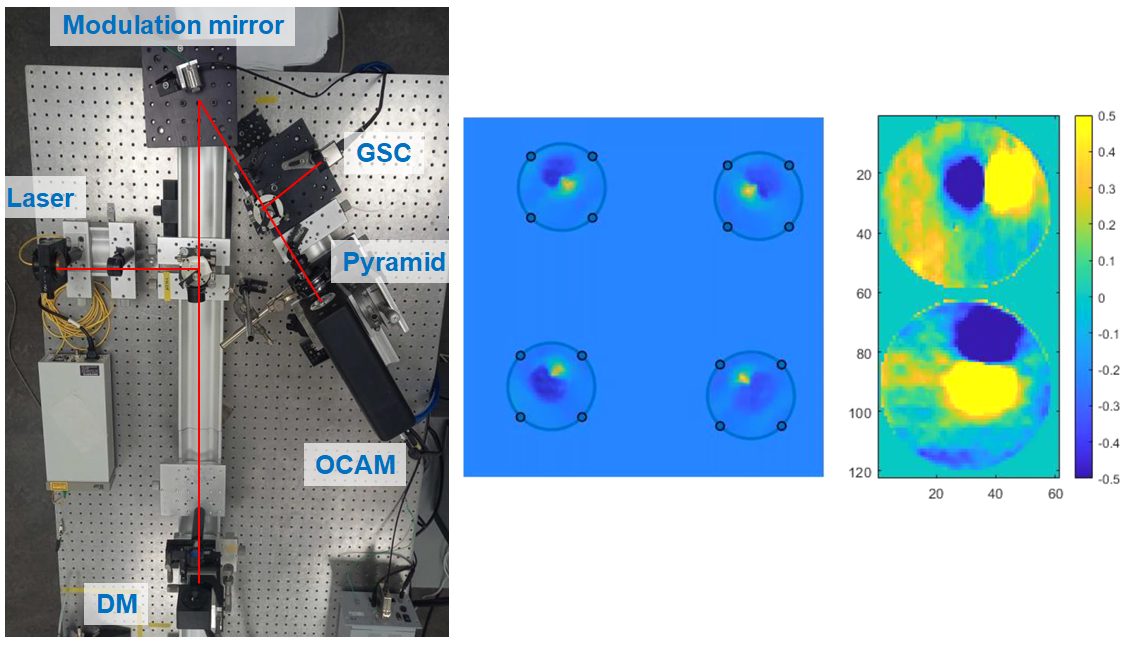}
    \caption{Tests of the WFS branch at LAM. Left: picture of the WFS branch. Middle: image seen on the WFS camera when a poke is sent to the DM. Right: computed slopes maps.}
    \label{fig:wfsbranch}
\end{figure}

The WFS branch is the core of the AO instrument. This branch has already been aligned at LAM (Fig. \ref{fig:wfsbranch}) to test the critical components to work together: the pyramid, the WFS camera, the DM and the modulation mirror. We processed the raw intensities and were able to compute the calibration matrix of the system by sending push-pull commands on the DM.\\

Once the tests performed, the WFS branch has been unmounted. The alignment of the full bench is currently under process at LAM.

\subsection{On-sky measurements at OHP}

Measurements at the Coude focus of the T152 telescope have been performed on July 20th and 21st, 2021. The objective of these measurements is to characterise the telescope pupil plane and focal plane. The possible high-amplitude movements of these planes is problematic for the AO system. However the pupil plane has shown stability better than $1\%$ of the diameter during a long-exposure observation. The movement of the pupil is consequently less than a fraction of point of measurement of the phase (equivalent to a sub-aperture for a Shack-Hartmann WFS).\\

The star position is targeted within a $\pm 2$ arcmin uncertainty diameter in the sky due to the telescope pointing lack of accuracy. However once the star is found it can be correctly centred manually. Then the PSF suffers a small drift of $\sim 20$ arcsec/hour. This drift is supposed to come from the telescope weight unbalance that might accelerate or slow down the Earth rotation compensation. However this drift is small and will be managed by the PAPYRUS slow tip-tilt mirror.

\begin{figure}[H]
\centering
	\includegraphics[width=0.7\columnwidth]{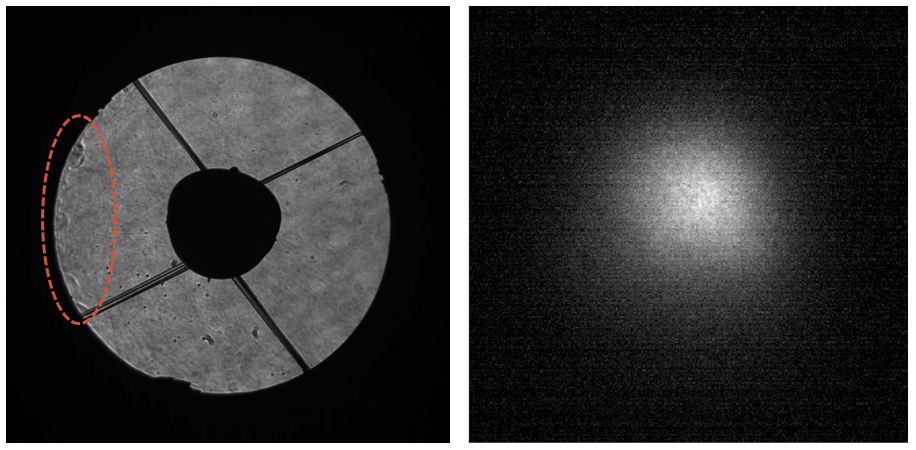}
    \caption{Characterisation of the OHP/T152 telescope. Left: image of the pupil. The dashed red ellipse shows presence of localised amplitude fluctuations at the edge of the pupil. Right: long-exposure PSF obtained on Vega.}
    \label{fig:T152test}
\end{figure}

\subsection{Schedule}

\begin{figure}[H]
\centering
	\includegraphics[width=0.9\columnwidth]{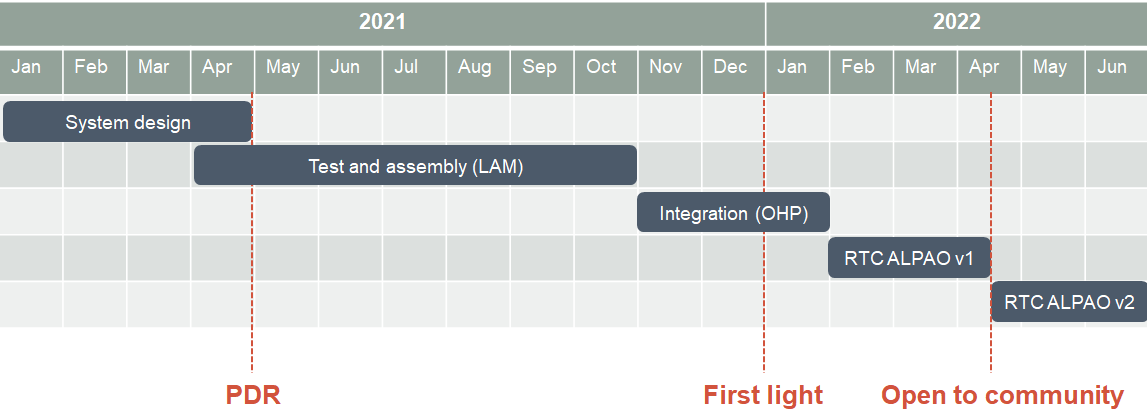}
    \caption{Schedule of the PAPYRUS project.}
    \label{fig:schedule}
\end{figure}

The schedule of the PAPYRUS project is given on Fig. \ref{fig:schedule}. Integration at OHP is planned for end 2021, and PAPYRUS will be available for the astronomical community beginning or mid 2022.

\section{Conclusion}

As it was presented here PAPYRUS will soon allow LAM to improved its operational expertise on Pyramids and deliver an AO corrected image at OHP. The system has been designed, the components have been characterised, and the first integration is underway with preparations for arrival at OHP also well advanced. The project will soon be opened to the community and we will welcome any new idea and collaboration to add new capabilities to the bench.

\section{Acknowledgement}

This project has been supported by ONERA, LAM and ALPAO in particular the WOLF ANR. We also thanks First Light and the OHP for their contribution this project.



\begin{thebibliography}{999}
\bibitem  {fauvarque2017} O. Fauvarque, B. Neichel, T. Fusco, J.-F. Sauvage, and O. Girault, “Generalformalism  for  fourier-based  wave  front  sensing:  application  to  the  pyramidwave  front  sensors,”Journal of Astronomical Telescopes, Instruments, andSystems3(1), p. 019001, 2017.
\bibitem {fauvarque2019kernel} O. Fauvarque, P. Janin-Potiron, C. Correia, Y. Brule, B. Neichel, V. Cham-bouleyron, J.-F. Sauvage, and T. Fusco, “Kernel formalism applied to fourier-based  wave-front  sensing  in  presence  of  residual  phases,”JOSA A36(7),pp. 1241–1251, 2019.
\bibitem {janin2019adaptive} P. Janin-Potiron, V. Chambouleyron, L. Schatz, O. Fauvarque, C. Z. Bond,Y. Abautret, E. R. Muslimov, K. El Hadi, J.-F. Sauvage, K. Dohlen,et al.,“Adaptive  optics  with  programmable  fourier-based  wavefront  sensors:   a spatial  light  modulator  approach  to  the  lam/onera  on-sky  pyramid  sensor testbed,”Journal of Astronomical Telescopes, Instruments, and Systems5(3),p. 039001, 2019.
\bibitem {chambouleyron2021focal} V. Chambouleyron, O. Fauvarque, J.-F. Sauvage, B. Neichel, and T. Fusco,“The focal-plane assisted pyramid wavefront sensor:  enabling frame-by-frame optical gains tracking,”arXiv preprint arXiv:2103.02297,2021.
\bibitem {fetick2020blind}  R. J. Fetick, L. Mugnier, T. Fusco, and B. Neichel, “Blind deconvolution in astronomy with adaptive optics: the parametric marginal approach,”MonthlyNotices of the Royal Astronomical Society496(4), pp. 4209–4220, 2020.
\bibitem {mugnier2004mistral} L.  M.  Mugnier,  T.  Fusco,  and  J.-M.  Conan,  “Mistral:   a  myopic  edge-preserving  image  restoration  method,   with  application  to  astronomical adaptive-optics-corrected long-exposure images,”JOSA A21(10), pp. 1841–1854, 2004.
\bibitem {OHPwebsite} OHP, “meteorological data.” http://www.obs-hp.fr/climatologie.shtml.
\bibitem {rigaut1998} F. J. Rigaut, J.-P. Veran, and O. Lai, “Analytical model for Shack-Hartmann-based  adaptive  optics  systems,”  in Adaptive Optical System Technologies,3353, pp. 1038–1048, International Society for Optics and Photonics, 1998.
\bibitem {conan2014} R. Conan and C. Correia, “Object-oriented matlab adaptive optics toolbox,”inAdaptive optics systems IV,9148,  p.  91486C,  International  Society  for Optics and Photonics, 2014.
\bibitem {CANON} Canon Inc.,“Canoncameramuseum.”https://global.canon/en/c-museum/product/ef434.html.
\bibitem  {Ohtake97} M.Ohtake and M. Mori, “Compact wide-angle objective lens,” Sept. 16 1997.US Patent 5,668,669A

\end{thebibliography}
\end{document}